\documentclass{article}
\usepackage[hyphens]{url}
\usepackage[T1]{fontenc} % add special characters (e.g., umlaute)
\usepackage[utf8]{inputenc} % set utf-8 as default input encoding
\usepackage{ismir,amsmath,cite,url}
\usepackage{graphicx}
\usepackage{color}
\usepackage{xcolor}
\usepackage{booktabs}
\usepackage[inline]{enumitem}

\usepackage{lineno}
\newcommand{\header}[1]{\textbf{#1}\hspace{0.75em}}

\newcommand{\mll}{ML\;}
\newcommand{\ai}{AI}

\newcommand{\eurovision}{pop\;}

\newcommand{\todo}[1]{\textcolor{black}{#1}}

\title{AI Song Contest: Human-AI Co-Creation in Songwriting}
\multauthor
{Cheng-Zhi Anna Huang$^1$ \hspace{1cm} 
Hendrik Vincent Koops$^2$ \hspace{1cm} Ed Newton-Rex$^3$} { \bfseries{Monica Dinculescu$^1$ \hspace{1cm} Carrie J. Cai$^1$}\\
 $^1$ Google Brain \hspace{1cm}
$^2$ RTL Netherlands \hspace{1cm}
$^3$  ByteDance\\
{\tt\small annahuang,noms,cjcai@google.com, h.v.koops,ednewtonrex@gmail.com}
}
\def\authorname{Cheng-Zhi Anna Huang, Hendrik Vincent Koops, Ed Newton-Rex, Monica Dinculescu, Carrie J. Cai}

\begin{document}

\maketitle
%
% [anna] abstract
% [anna] intro

% [vincent, anna] contribution

% [anna] related work
% [vincent] approach
% [anna] high-level co-creation approaches

% [vincent] workflow
% [vincent] impact on creativity, experience of co-creation

% [vincent] strategies
% [anna] use of AI merge with Vincent
% - manual versus feedback...
% short description
% high-level
% low-level (from AI perspective)
% end with discussion recommendation
% [anna] accessibility / usability
% [vincent] ethics consideration

% [anna] judging AI songs
% [] final discussion
% [] conclusion

% question: 
% fix flow from "setting ideas into musical context" to "establish social context and personal agency"
%
%

% \paragraph{Notation}
% \begin{itemize}
%     \item One team did something ... (T13)
%     \item Multiple teams did something ... (T13, T12) (T13,12,8,9)
% \end{itemize}

%The abstract should be placed at the top left column and should contain about 150-200 words.
\begin{abstract}
Machine learning is challenging the way we make music. Although research in deep generative models has dramatically improved the capability and fluency of music models, recent work has shown that it can be challenging for humans to partner with this new class of algorithms. In this paper, we present findings on what 13 musician/developer teams, a total of 61 users, needed when co-creating a song with AI, the challenges they faced, and how they leveraged and repurposed existing characteristics of AI to overcome some of these challenges. Many teams adopted modular approaches, such as independently running multiple smaller models that align with the musical building blocks of a song, before re-combining their results. As ML models are not easily steerable, teams also generated massive numbers of samples and curated them post-hoc, or used a range of strategies to direct the generation, or algorithmically ranked the samples. Ultimately, teams not only had to manage the ``flare and focus'' aspects of the creative process, but also juggle them with a parallel process of exploring and curating multiple ML models and outputs. These findings reflect a need to design machine learning-powered music interfaces that are more decomposable, steerable, interpretable, and adaptive, which in return will enable artists to more effectively explore how AI can extend their personal expression.  

\end{abstract}
\section{Introduction}\label{sec:introduction}
% \newpage
%Todos: 
% - find quote for yacht, and also reference
% - check out bob sturm's paper and see what his framing is
% also dorien, chuan and chew's survey paper that is structured by musical components
%
% Background: 
% electronic sequencers
% Maybe have musicians be the actor
%
%Technology has always helped expand the range of musical expression, from the fortepiano to synthesizers to drum machines. Can machine learning further extend our creative palette? 
%
% ask the question of shifting dynamics of AI and artists? perhaps after the examples.
% at the end come back to how we want to build richer conditioning / interaction, so that artists still use them as a tool even when the technology improves tremendously.
% extend human creativity? 
% first give examples of prominent use that embodies "co-creation", then introduce the term co-creation...
% - how do we "define" co-creation? 
% Use musicians or songwriters?
Songwriters are increasingly experimenting with machine learning as a way to extend their personal expression~\cite{barbican2020songs}.
%There's been an increasing number of musicians experimenting with using \ml asMusicians have been using machine learning in highly personal ways~\cite{barbican2020songs}
%symbolic, Yacht~\cite{}, 
% TODO: check if dance punk is appropriate for yacht
For example, in the symbolic domain, the band Yacht used MusicVAE~\cite{roberts2018hierarchical}, a variational autoencoder type of neural network, to find melodies hidden in between songs by interpolating between the musical lines in their back catalog, commenting that \emph{``it took risks maybe we aren’t willing to''}~\cite{mattise2019yacht}. 
% YACHT singer Claire Evans says that the process allowed the band to find melodies "hidden in between songs" in their back catalog
% "I don’t know if we could’ve written it ourselves—it took a risk maybe we aren’t willing to."
% TODO: for DISCUSSION
% Yacht in between... there existing melodies that can actually be awkward to play/has its own groove
%
% audio, Holly Herndon ~\cite{maas2020holly} 
In the audio domain, Holly Herndon uses neural networks trained on her own voice to produce a polyphonic choir.~\cite{maas2020holly,denton2019holly}.
In large part, these human-AI experiences were enabled by major advances in machine learning and deep generative models~\cite{oord2016wavenet, vaswani2017attention, dieleman2018challenge}, many of which can now generate coherent pieces of long-form music~\cite{huang2019music, hawthorne2018enabling, payne2019muse, dhariwal2020jukebox}.
%~\cite{oord2016wavenet, dieleman2018challenge, huang2019music, hawthorne2018enabling, payne2019muse, dhariwal2020jukebox}. 

Although substantial research has focused on improving the algorithmic performance of these models, much less is known about what musicians actually need when songwriting with these sophisticated models. 
%Research suggests that, 
Even when composing short, two-bar counterpoints, it can be challenging for novice musicians to partner with a deep generative model: users desire greater agency, control, and sense of authorship vis-a-vis the AI during co-creation~\cite{louie2020novice}.

Recently, the dramatic diversification and proliferation of these models have opened up the possibility of leveraging a much wider range of model options, for the potential creation of more complex, multi-domain, and longer-form musical pieces. Beyond using a single model trained on music within a single genre, how might humans co-create with an open-ended set of deep generative models, in a complex task setting such as songwriting? 

In this paper, we conduct an in-depth 
% examination 
study
of what people need when partnering with AI to make a song. Aside from the broad appeal and universal nature of songs, songwriting is a particularly compelling lens through which to study human-AI co-creation, because it typically involves creating and interleaving music in multiple mediums, including text (lyrics), music (melody, harmony, etc), and audio. This unique conglomeration of moving parts introduces unique challenges surrounding human-AI co-creation that are worthy of deeper investigation. 

As a probe to understand and identify human-AI co-creation needs, 
% i changed this for anonymity and also since we didn't organize the contest, and only served as judges 
%we conducted a large-scale human-AI songwriting contest 
we conducted a survey during a large-scale human-AI songwriting contest, in which 13 teams (61 participants) with mixed (musician-developer) skill sets were invited to create a 3-minute song, using whatever AI algorithms they preferred. Through an in-depth analysis of survey results, we present findings on what users needed when co-creating a song using \ai, what challenges they faced when songwriting with \ai, and what strategies they leveraged to overcome some of these challenges. 

We discovered that, rather than using large, end-to-end models, teams almost always resorted to breaking down their musical goals into smaller components, leveraging a wide combination of smaller generative models and re-combining them in complex ways to achieve their creative goals. Although some teams engaged in active co-creation with the model, many leveraged a more extreme, multi-stage approach of first generating a voluminous quantity of musical snippets from models, before painstakingly curating them manually post-hoc. Ultimately, use of AI substantially changed how users iterate during the creative process, imposing a myriad of additional model-centric iteration loops and side tasks that needed to be executed alongside the creative process. Finally, we contribute recommendations for future AI music techniques to better place them in the music co-creativity context.

In sum, this paper makes the following contributions:
% \todo{
% %\begin{enumerate*}
% %\item{Results from 13 teams (61 people) co-creating songs with AI.}
% \item{A description of common patterns these teams used when songwriting with a diverse, open-ended set of deep generative models.}
% \item{An analysis of the key challenges people faced when attempting to express their songwriting goals through AI, and the strategies they used in an attempt to circumvent these AI limitations.}
% \item{Implications and recommendations for how to better design human-AI systems to empower users when songwriting with AI.}
% \end{enumerate*}
% }
\vspace{-5pt}
\begin{itemize}
%\item{Results from 13 teams (61 people) co-creating songs with AI.}
\item{A description of common patterns these teams used when songwriting with a diverse, open-ended set of deep generative models.}
\vspace{-4pt}
\item{An analysis of the key challenges people faced when attempting to express their songwriting goals through AI, and the strategies they used in an attempt to circumvent these AI limitations.}
\vspace{-4pt}
\item{Implications and recommendations for how to better design human-AI systems to empower users when songwriting with AI.}
\end{itemize}

\section{Related work}
Recent advances in AI, especially in deep generative models, have renewed interested in how AI can support mixed-initiative creative interfaces~\cite{mici2017deterding} to fuel human-AI co-creation~\cite{geyer2020hai}.  \emph{Mixed initiative}~\cite{horvitz1999principles} means designing interfaces where a human and an AI system can each ``take the initiative'' in making decisions. \emph{Co-creative}~\cite{liapis2016mixedinitiative} in this context means humans and AI working in partnership to produce novel content. For example, an AI might add to a user's drawing ~\cite{fan2019collabdraw,oh2018lead}, alternate writing sentences of a story with a human~\cite{clark2018creative}, or auto-complete missing parts of a user's music composition~\cite{bazin2019nonoto,huang2018mixed,huang2019bach,louie2020novice}. 
Within the music domain, there has been a long history of using AI techniques to model music composition~\cite{papadopoulos1999ai, fernandez2013ai,pasquier2016introduction, herremans2017functional}, by assisting in composing counterpoint~\cite{farbood2001analysis, herremans2013composing}, harmonizing melodies~\cite{pachet2001musical, chuan2007hybrid,koops2013functional}, more general infilling~\cite{hadjeres2016deepbach,huang2017coconet,ippolito2018infilling,shaw2019multitask}, exploring more adventurous chord progressions~\cite{nichols2009data, fukayama2013chord, huang2016chordripple},
%adjusting the mood of a harmonization
semantic controls to music and sound generation~\cite{simon2008mysong,cartwright2013social,huang2014active, ferreira2019learning}, building new instruments through custom mappings or unsupervised learning~\cite{fiebrink2011real, donahue2019piano}, and enabling experimentation in all facets of symbolic and acoustic manipulation of the musical score and sound~\cite{agon2006om, bresson2008om}. 

More recently, a proliferation of modern deep learning techniques~\cite{briot2017deep, briot2020artificial}  
has enabled models capable of generating full scores~\cite{huang2020pop}, or producing music that is coherent to both local and distant regions of music~\cite{huang2019music,payne2019muse}.
%or producing music that complements both local and distant regions of music \todo{cite}. 
The popular song form has also been an active area of research to tackle modeling challenges such as hierarchical and multi-track generation~\cite{ papadopoulos2016assisted, roberts2017hierarchical, simon2018learning, zhou2018bandnet}. 

Despite significant progress in deep generative models for music-making, there has been relatively little research examining how humans interact with this new class of algorithms during co-creation. A recent study on this topic~\cite{louie2020novice} found that deep learning model output can feel non-deterministic to end-users, making it difficult for users to steer the AI and express their creative goals.  Recent work has also found that users desire to retain a certain level of creative freedom when composing music with AI~\cite{frid2020music,louie2020novice,sturm2019machine}, and that semantically meaningful controls can significantly increase human sense of agency and creative authorship when co-creating with AI~\cite{louie2020novice}. While much of prior work examines human needs in the context of co-creating with a single tool, we expand on this emerging body of literature by investigating how people assemble a broad and open-ended set of real-world models, data sets, and technology when songwriting with AI.

\section{Method and Participants}

\todo{The research was conducted during the first three months of 2020, at the AI Song Contest organized by \textsc{VPRO}~\cite{dijk2020ai}. The contest was announced at ISMIR in November 2019, with an open call for participation.} Teams were invited to create songs using any artificial intelligence technology of their choice. The songs were required to be under 3 minutes long, with the final output being an audio file of the song.
At the end, participants reflected on their experience by completing a survey.
Researchers obtained consent from teams to use the survey data in publications. 
\todo{The survey consisted of questions to probe how teams used AI in their creative process:}
\vspace{-4pt}
\todo{
\begin{itemize}
%\item{How did teams use AI in their creative process?}
\item{How did teams decide which aspects of the song to use AI and which to be composed by musicians by hand? What were the trade-offs?}
\vspace{-4pt}
\item{How did teams develop their AI system? How did teams incorporate their AI system into their workflow and generated material into their song?}
%\item{What were some of the challenges and trade-offs?}
\end{itemize}
}

% The research was conducted during the first three months of 2020, at a contest where teams were invited to create songs using any artificial intelligence technology of their choice. The songs were required to be under 3 minutes long, with the final output being an audio file of the song. At the end, participants reflected on their experience by completing a survey. Researchers obtained consent from teams to use the survey data in publications. The survey consisted of questions about:
% %1) How did they use AI in their creative process? and 2) How did AI expand or limit their creativity?
% \todo{
% \begin{itemize}
% \item{How did teams use AI in their creative process?}
% \item{How did teams decide which aspects of the song to use AI and which to be composed by musicians by hand?}
% \item{How did teams approach develop their AI system?} 
% \item{How did teams incorporate their AI system into their workflow? How did teams work the generated material into their song?}
% \item{What were some of the challenges and trade-offs?}
% \end{itemize}
% }
%
%
%\todo{From the 25 registrations, 
\vspace{-4pt}
In total, 13 teams (61 people) participated in the study. The teams ranged in size from 2 to 15 (median=4). 
\todo{Nearly three fourths of the teams had 1 to 2 experienced musicians.}
% 10 out of 13 teams had experienced musicians
\todo{A majority of teams had members with a dual background in music and technology: 5 teams had 3 to 6 members each with this background, and 3 teams had 1 to 2 members.} 
% 8 out of 13 had folks with mixed background, 5 teams had 3 to 6 members with dual background
%and consisted of a mix of musicians and developers, many of whom had a dual background in music and technology. 
%Those who were musicians often had professional-level songwriting and production experience. 
%Participants were recruited through a study announcement. 
We conducted an inductive thematic analysis~\cite{braun2006using,braun2014can,adams2008qualititative} on the survey results to identify and better understand patterns and themes found in the teams' survey responses. One researcher reviewed the survey data, identifying important sections of text, and two researchers collaboratively examined relationships between these sections, iteratively converging on a set of themes.

\begin{table*}[]
\centering 
\resizebox{2\columnwidth}{!}{%
\begin{tabular}{ll}
\toprule
 Music building blocks & Models \& techniques   \\
%   \hline%
\midrule
\noalign{\vskip .5ex}
% i'm commenting out LSTM-GAN b/c team 10's result is not as successful as a song
%Lyrics               & LSTM-GAN, GPT-2, LSTM, Transformer                                                                                   \\
Lyrics               & GPT2, LSTM, Transformer                                                                                   \\
\midrule
% i'm commenting out vocal melody b/c it was just added in thumbnails to show where the artist added their own vocals
%Vocal melody         & LSTM-GAN, VAE, WaveNet + LSTM, SampleRNN, GAN, Transformer                                                           \\
%Melody               & VAE, WaveNet + LSTM, Char-RNN, Sample-RNN, GAN, Transformer, CNN, custom "MiniVAE", LSTM-GAN, MusicVAE, MusicAutobot \\
Melody               & CharRNN, SampleRNN, LSTM + CNN, WaveNet + LSTM, GAN, Markov model  \\ 
 %                   & MiniVAE, MusicVAE, MusicAutobot \\
% Harmony              & LSTM, RNN auto-encoder, Transformer, GAN, MusicAutobot, Coconet/Coucou                                                      \\

Harmony              & LSTM, RNN autoencoder, GAN, Markov model \\
% Bassline             & LSTM+CNN, WaveNet+LSTM, Transformer, GAN, MiniVAE, MusicVAE                                 \\
Bassline             & LSTM + CNN, WaveNet + LSTM, GAN \\
%Drums                & VAE, "Magenta midi generator", custom "MiniVAE" trio, Neural Drum Machine, Sample-RNN                               \\
% Drums                & DrumRNN, Neural Drum Machine, SampleRNN, MiniVAE trio                              \\
Drums                & DrumRNN, Neural Drum Machine, SampleRNN, Markov model \\
Multi-part     & MusicVAE trio (melody, bass, drums), MiniVAE trio, Coconet/Coucou (4-part counterpoint), \\
                & MusicAutobot (melody, accompaniment), Transformer (full arrangement) \\
Structure          & Markov model                                                                                             \\
\midrule
Vocal synthesis      & WaveNet, SampleRNN, Vocaloid, Sinsy, Mellotron, Emvoice, Vocaloid, custom vocal assistant                                                  \\
Instrument synthesis & SampleRNN, WaveGAN, DDSP                          \\
 \bottomrule     
\end{tabular}%
}
\caption{Overview of musical building blocks used by teams.}\label{tab:bblocks}
\end{table*}

\section{How did teams co-create with \ai?}

The vast majority of teams broke down song composition into smaller modules, using multiple smaller models that align with the musical building blocks of a song, before combining their results: \textit{``So my workflow was to build the song, section by section, instrument by instrument, to assemble the generated notes within each section to form a coherent whole''} (T12). A few teams first attempted to use end-to-end models to generate the whole song at once, such as through adversarial learning from a corpus of pop song audio files (T6) or through generating an audio track using SampleRNN~\cite{mehri2016samplernn} (T13). However, they quickly learned that they were unable to control the model or produce professional quality songs, and thus turned to the modular approach instead. \todo{In the following sections, we summarize how teams used modular building blocks, combined and curated them, and in some cases more actively co-created with AI to iterate on the outcomes.}

% I was trying to save space and removed one level of subsections "leveraging musical building blocks", and bringing subsubsections "creating musical building blocks" and "combining building blocks" up to subsections, and changing the first section title from "creating musical building blocks" to "leveraging modular musical building blocks".
\subsection{\todo{Leveraging modular musical building blocks}}
% \todo{tell one or two interesting stories of how/why they broke it down, with quotes}\todo{quote for why or how teams preferred a modular approach} 
%For example, one team tried to directly generate a song via adversarial learning from a corpus of pop song audio files. However, the team realized this was too ambitious, and resorted to \textit{``more traditional software audio synthesis.''} (T6) 
%``Hence, we limited our ambitions to generate midi files in the same way and generate the final audio signal with more traditional software audio synthesis.'' (T6).
%They imagined their outcome to be a neural network (or an ensemble of them) that directly outputs a synthesized song. 
%Similarly, a different team attempted to use SampleRNN to directly generate an audio track. After generating 10 hours of material, the team realized the material wasn't strong enough on its own. \emph{``Instead of writing code that writes a song, we decided to free up that requirement. We would approach it as music producers using AI in our
%creative workflow to come up with a killer track!''}, adding that this approach was successful, and \emph{''It was more fun this way too.''} Instead of end-to-end, the eventual song was created in a modular fashion, using separate models for different musical components. The failed SampleRNN output however did serve a purpose: as inspiration for the lyrics, storytelling and melody of the eventual piece (T13). 

\begin{figure}[h!]
 \centerline{
 \includegraphics[width=\columnwidth, trim={0.5cm 0.5cm 0.5cm 0.5cm},clip]{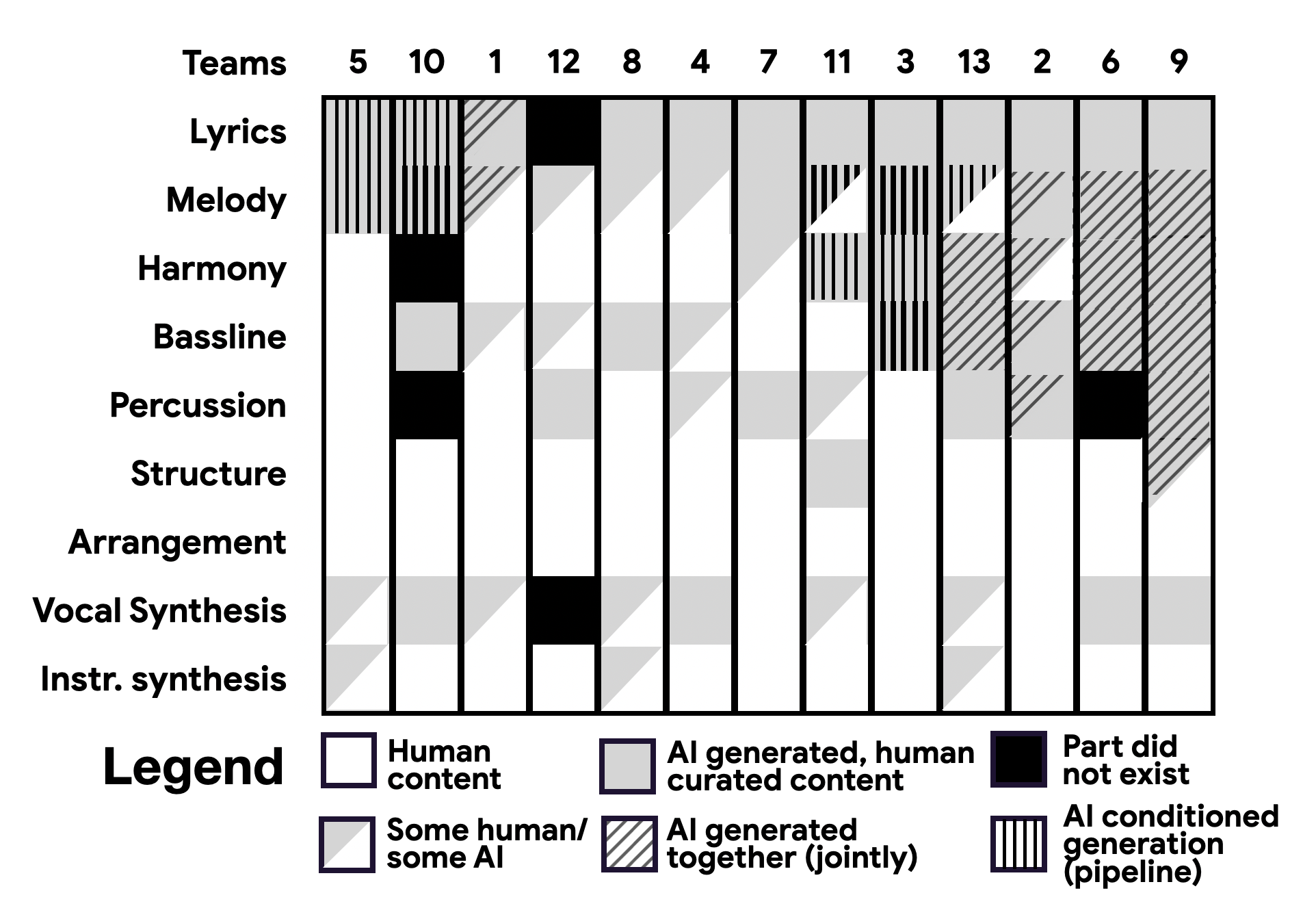}}
 \caption{\todo{An overview of how 13 teams co-created with AI in songwriting. %Each team’s song is illustrated as a column indicating whether each musical component is musician composed, AI generated then human curated, or both. 
 Each column shows whether each song's component was musician composed, AI generated  then human curated, or both.
 Nearly all teams manually aligned AI generated \emph{lyrics} and \emph{melody}, except teams in the first three columns. T5 used a two-stage pipeline by first generating \emph{lyrics} and \emph{melody} separately, then algorithmically matching them up using their stress patterns. T10 first generated \emph{lyrics}, and then conditioned on \emph{lyrics} to generate \emph{melody}. T1 jointly modeled \emph{lyrics} and \emph{melody}, generating syllables and notes in an interleaving fashion. 
 %The remaining columns are grouped by how teams combined other building blocks.
 T12, 8, 4, 7 all generated \emph{melodic lines} first, and then manually \textbf{stitched} them together by layering them vertically as \emph{melody} and \emph{bassline} to yield \emph{harmony}. In contrast, T11, 3 first generated \emph{chords}, then conditioned on \emph{chords} to generate \emph{melody} (and \emph{bassline} separately) in a \textbf{pipeline}.  T13 iterated between conditioning on \emph{melody} to jointly generate the other parts and vice versa. T2, 6 and 9 focused on models that could \textbf{jointly} generate multiple parts at once.}}
 % T2, 6 and 9 focused on multi-part models because they could \textbf{jointly} generate melody, harmony, bass, drums together in a coherent way, providing them with larger building blocks to work with.}}
\label{fig:thumbnails}
\end{figure}

Overall, teams leveraged a wide range of models for musical components such as lyrics, (vocal) melody, harmony, bassline, drums, arrangement, and vocal and instrument synthesis. Table \ref{tab:bblocks} shows an overview of models used for each song component, and Figure \ref{fig:thumbnails} illustrates how the 13 teams co-created with AI along these different components. 
%The number of unique models used by teams ranged from 1 to 5 (median 4). 
The number of unique model types used by teams ranged from 1 to 6 (median 4).
Some teams used the same type of model for modeling different song components.
%Some teams used the same type of model for generating multiple song components (e.g. melody, chords).

%  The remaining columns are grouped by how teams combine other building blocks, by generating each separately and then manually \textbf{stitching} them together (T12, 8, 4, 7), in a \textbf{pipeline} fashion where one part is generated (thicker vertical stripes) conditioning on another part (thinner vertical stripes, T11, 3, 13) or generating multiple parts \textbf{jointly} together (diagonal stripes, T13, 2, 6, 9).

All teams used AI to generate lyrics and melodies, and more than half of the teams synthesized vocals for parts of the song. 
Some of these decisions were due to model availability and existing levels of model performance. For example, teams almost always generated lyrics using AI because \todo{high-performing models like GPT-2~\cite{radford2019language} along with a fine-tuning script were readily available}.
%, along with a fine-tuning script. 
%
% %\todo{MORE STUFF ON WHY}
% \todo{trying to explain the way in combining building block sections}
%Teams with professional music experience often chose to only generate melodic lines (in addition to lyrics), in order to leave the artist enough creative freedom to compose the harmony and drums, arrange the piece, and decide how the various lines can be put together. 

Teams with professional musicians often chose to only generate lyrics and melodic lines, in order to \todo{leave enough creative space to musicians to decide how the various lines can be put together  
%compose harmony, percussion, 
and to arrange the song in their own style (T5, 8). One exception is T3 who generated a lead sheet with lyrics, melody and harmony.}
%and decide how the various lines can be put together. 
%Hence these lower parts were less frequently generated. One exception is T3 who generated lead sheets of lyrics, melody and harmony.

Teams with more ML and less music expertise opt for minimal arrangements (T10, 6, 9), and often used multi-part models because they could generate melody, harmony, bass, drums together in a coherent way, providing teams with larger building blocks to work with. In one extreme case, \todo{the team was able to generate all sections of their song by traversing the latent space of MusicVAE (T9) (see ``Bridging sections through latents" below for more detail)}.
\subsection{Combining building blocks}
% Outline
% - Cascading pipeline
% - Human has lyrics, melody, etc., human puts it together
%   -- Fitting melody to the lyrics
%   -- Stress patterns
%   -- Human creates the right # of notes given lyrics

% \todo{horizontal versus vertical}
% \todo{show some different orderings!
% Like lyrics then melody then chords
% Or chords then melody then lyrics
% Or melody then lyrics then chord}

%An example for changing a model after exploration is a team that 
Teams leveraged many strategies for combining outputs of smaller models, piecing together the musical building blocks to form a coherent song.
These ranged from manually combining individual components, to using heuristics to automatically pair them up,
to creating a pipeline between models, to adopting models that jointly model multiple components at once. 

\header{Stitching} 
%\todo{As none of the teams used an end-to-end approach, all teams manually ``stitched'' together machine generated material.} 
\todo{Many teams manually ``stitched'' together machine generated material, with the result informing the manual composition of other musical components.}
%\todo{T1 described their songwriting process as manually \textit{``curating material generated by an AI and then assembling it together to form a coherent whole''}.}
% T1 “the song would be created by curating material generated by an AI and then assembling it gotether to form a coherent whole”.
% Similarly,
\todo{In one team, a musician \textit{``selected, tweaked, and weaved AI-generated melody, chords and drum parts into a ballad song form''}, while another musician wrote the bassline \textit{``that seemed more or less implied by the AI materials''} (T7).}
%T7, "Nancy tweaked the AI-generated melody, chords and drum parts into a ballad, while Collins wrote the bassline, which seemed more or less implied by the AI materials. The text was also created by AI, but the combination, instrumentation and production of these materials comes from Collins and Carlisle."
%"Tom wrote the bass line because he felt like it was more-or-less implied by the melodies and chords that Nancy had selected, and arranged the song in a digital audio workstation."
%"selected, tweaked, and weaved this into a recognisable and beautiful song form.
\todo{This is echoed by another team, who composed the accompaniment \textit{``based on chordal movements predicted by the melodic fragments''} (T5).}  
%T5, manual accompaniment, “based on chordal movements predicted by the melodic fragments”.
%
%Multiple teams manually ``stitched" together machine generated material, by layering multiple melodic lines on top of each other to form harmony (T1, 12, 8, 4). 
\todo{Several teams layered melodic lines to yield harmony (T1, 12, 8, 4)}.

%\todo{human creates right number of notes given lyrics}
% Moved all examples of lyrics to the last section. 
%Others leveraged automatic heuristics to scaffold this stitching process. 
%

% \todo{pipelining}
% Or if multiple musical components are composed by AI, it becomes a pipeline
% T8, first generated rhythm, “when notes should be played within a measure”,  then fill in the melody and indicate if a chord was to be played, and then another network for deciding on the chord

\header{Pipelines} Several teams leveraged model pipe-lining, feeding the output of one model into the input of another model. 
\todo{To generate melody that aligns well with lyrics, one team first used GPT-2 to generate lyrics, then a lyric-conditioned model to generate melody~\cite{yu2019conditional} (T10).}
%\todo{Another team generated a melody from GPT-2 generated lyrics (T10).}
% example mentioned in figure and also two stage curation
%\todo{In contrast, another team first generated lyrics and melody separately, and then \emph{``filtered''} them algorithmically by pairing up lyrics and melody using their stress patterns (T5).}
One team decomposed melody generation into two steps, first using a LSTM to generate rhythm as note onset patterns, and then a CNN to assign a pitch to each onset (T8).  
%
%They also experimented with
%after which a CNN melody-network fills in the onsets with pitches. Extending the pipeline, the melody network will also determine if a chord should be played, in which case a separate LSTM chord-network finds an appropriate chord. 
%Another team example of pipelining is Lyrics-Conditioned Neural Melody Generation to generate a melody from GPT-2 generated lyrics (T10).
%
%
%
\todo{While many teams ``stitched'' together melodic lines to create harmony, two teams first generated chords and then melody (and bassline) (T11 and T13)}. %
%T3 used a LSTM autoencoder to generate chords, and then another LSTM to condition on those chords to generate melody. T11 used a similar pipeline but first LSTM for chords, and then used statistics on note duration and intervals, along with the chords to generate melody??.} 
%
Pipeline approaches allowed teams to refine the output at each intermediate stage before passing content into the next model.
%
%some teams compose chords first and then conditioned on chords to generate melodic lines, while other teams first composed melodic lines, and then stitched them together to yield harmony
%
%\todo{combining lyrics} using algorithmic or model-based approaches
%

\header{Joint modeling} To generate multiple parts together, several teams adopted models such as MusicVAE trio~\cite{roberts2017hierarchical}, Coconet~\cite{huang2017coconet}, MusicAutobot~\cite{shaw2019music} or Transformers that are trained to jointly model multiple instrumental parts \todo{(T13, 2, 6, 9)}. 
%A few teams tried using joint modeling of multiple instruments to algorithmically combine musical components. 
%(T1) In a unique approach, one team built a model capable of concurrently generating melody and lyrics (T1). Using this model, 100 melody-lyrics sequences were generated, from which cherry-picked elements were used to compose a song. 
One team experimented with jointly modeling notes and syllables from pairs of melodies and lyrics, %by representing lyrics in a syllabic way, 
but found it \textit{``very hard to concurrently generate semantically meaningful lyrics and a melody that is not aimless''} (T1).

\subsection{Generate then curate}
%\vspace{-3pt}
A common approach was to generate a large quantity of musical samples, followed by automatically or manually curating them post-hoc. Teams took a range of approaches to curating the large quantity of results, ranging from brute-force manual curation, to a two-stage process of first filtering with AI, then curating manually.

\header{Generation} Often, teams used models to generate a large volume of samples to choose from. For instance, one team used their pipeline LSTM + CNN model to generate over 450 melodies and basslines (T8). Another team generated 10K lines of death metal lyrics (T13). 
%lyrics in the style of death metal: \emph{``It generated over 10k lines of lyrics very dismal, brutal, and gorey (sic) \textellipsis To our surprise, many of them had to do with infections and disease. There was no going back.''} (T13). 

\header{Manual curation} While curating, teams were often looking for the key musical themes or motifs to use for their song. For example, one team used MusicVAE to generate several combinations of lead, bass, and drums, and \textit{``handpicked the most appealing''} version to serve as their verse (T2).  
%Another team generated chord sequences using an RNN, from which a sequence was chosen that the team was happy with (T3). 
Another team was looking for a small, catchy snippet or \textit{``earworms"} to flesh out the music (T11).
%Some teams identified ``earworms'', or catchy snippets from model outputs, through manual inspection and selection: \textit{``we had [the model] generate a bulk of 100 sequences and then found the most interesting material by listening to them."} 
%One team (T3) wanted to change the sentiment of the generated lyrics to create a positive song, and decided to \emph{``filter out all negative songs from the dataset, by analyzing the sentiments of each song''}.

\header{Two-stage curation} 
A few teams first used automated methods to narrow down the choices, before manually selecting what would fit best in their song. 
For example, one team used a \textit{``catchiness''} classifier trained on rankings of songs to filter melodies before handing them to an artist (T8).  %\todo{<-- need to distinguish between quotes that are user quotes vs. Anna quotes} 
%For example, one team estimated the \emph{catchiness} and \emph{eurovisionness} of the lead and bass melodies using the FANTASTIC feature toolbox and a LASSO regression model \cite{mullensiefen2009fantastic} (T8). They also used musical transformations to analyse the generated and original pieces from to filter out material that was too similar to original 
% ESC 
%melodies \cite{melkonian2019constitutes}. 
%
Another team curated their generated material by first algorithmically matching up the stress patterns of lyrics and melodies to make sure the material could be more immediately useful (T5).
%, before further manual refinement by musicians (T5). 

Several teams found the process of generating and curating painstaking, or similar to a difficult puzzle (T1). However, one team described this massive generation process as exhilarating, or like \textit{``raging with the machines"} (T5). They most appreciated the unexpected surprises, and actively engaged with this firehose of raw, AI-generated material: \textit{``We couldn't resist including as much of the good and quirky machine output as possible...makes it much less repetitive than much of the music we might produce as humans. We really enjoyed having this firehose of generative capability...its constantly moving nature"} (T5). 
%something that is really beguiling and engaging in 
%, more usable content than we'd usually have to deal with in a track, making

%: \emph{``What I love most about death metal and its cookie monster vocals is that sometimes it’s so absurdly funny. GPT2 lines like \emph{`Lying dead on the floor. No head rest assured'} had me deep in the lmaos''}. (T3)

% \todo{TODO: this paragraph that the teams curated, but need to explain *how* they curated -- how did they sift through this extremely large volume of stuff??} \todo{more details on the process  and quotes added}

\subsection{Active co-creation}
%Four teams
Some teams co-created music with AI in a more blended manner, where the model outputs influenced human composing and vice versa, similar to how human musicians might work together to write a song.

%\header{Singing along} 
A few teams used AI-generated output as an underlying foundation for composing on top of, such as improvising a melody over AI-generated chords: \textit{``we played the chords, and all of us around the table hummed along until we got to a simple and catchy melody''} (T11). Others took the AI output as raw material generated in a predefined structure, and manually composed an underlying beat (T8). 

Others took AI output as an initial seed for inspiration and further elaboration. For example, one team trained SampleRNN on %1950s 
 acappellas, which generated nonsensical output similar to babbling. A musician then tried to 
``transcribe'' the words and the melody, and sang along with it.
%recognize words in this output and applied it to the main singing voice: 
\textit{``She found sections that sounded like lyrics. She wrote those lyrics down and sang them. She spent a day riffing on those lyrics, building a dialogue"}. 
These riffs and words 
%eventually
fueled the formation of the larger story and song (T13). 

%\header{Jamming} 
For one participant, working with the AI was like jamming with another musician, an active back and forth process of priming the AI with chord progressions, hearing AI output, riffing on that output, then using those new ideas to seed the AI again (T12). 
% \todo{ANNA: I tried to clarify this last user's process, please confirm this is actually right}. 

%he would communicate the chord progressions he had in mind through the notes he was playing, see what the AI comes up with, and then play along or try something else.
%They argued that melody generation and lyric-to-melody techniques are \emph{``challenging research subjects and we did not have the time to delve in''}. (T11).

One team described making some deliberate decisions about how much agency to provide the AI vs. themselves as artists. To preserve the AI's content, an artist tried to only transpose and not ``mute'' any of the notes in two-bar AI-generated sequences, as he chose which ones to stitch together to align with lyrics. However, to bring the artist's own signature as a rapper, he decided to compose his own beat, and also improvise on top of the given melodies freely with a two-syllable word that was made up by \mll (T8).

\section{Teams overcoming AI limitations}

% \subsection{Adapting existing models for a different style}
% (often with a much smaller dataset)
% Fine-tuning pretrained checkpoints (i.e. two stage fine-tuning on song lyrics and then Eurovision song lyrics for more uplifting lyrics)
% Check for overfitting
% Training mini VAEs on top of the latent space of musicVAE (i.e. MidiMe) [T9]
% Training from scratch [T7]

In the previous section, we described the ways in which participants co-created with AI. Although teams made some headway by breaking down the composition process into smaller models and building blocks, we observed a wide range of deeper challenges when they attempted to control the co-creation process using this plethora of models. In this section, we describe participants' creative coping strategies given these challenges, and the strategies they used to better direct the co-creation process.

\subsection{\mll is not directly steerable}
Due to the stochastic nature of \mll models, 
%algorithmic output can be unpredictable. 
their output can be unpredictable. 
During song creation, teams leveraged a wide range of strategies in an attempt to influence model output, 
% such as through data during training or fine-tuning, 
such as through data during fine-tuning, 
or through input or conditioning signals during generation. Below, we describe the most common patterns observed:

% \header{Training}
% % \todo{I really like this example, but I'm not sure where to place it under. if folkrnn trained from pretrained model then finetuning. Checking. Apparently they trained it from scratchm, so no fine-tuning}
% To see whether they could generate a 
% % Eurovision 
% song with a Dutch twist, a team used different mixtures of popular music
% % ESC 
% and Dutch folk songs to train FolkRNN. The team reflected on the output and realized that it did not lead to a new and coherent style-mix, but rather interesting melodies of which some segments were of one style, while other segments of another style. The team also noted that adding the Dutch folk songs helped against overfitting on 
% their
% % the ESC 
% dataset (T8). 

\header{Fine-tuning}
%For example, teams fine-tuned and retrained GPT-2 to produce lyrics that were more \todo{what type of lyrics}. 
%For \nounheaders{lyrics}, \noun{GPT-2} was often fine-tuned with lyrics datasets, 
After experimenting with a model, many teams tried to influence the mood or style of the generated content by fine-tuning models on a smaller dataset. For example, teams fine-tuned GPT-2 with lyrics from uplifting \eurovision songs to German death metal (T13), in order to steer the generation towards phrase structures more common in lyrics and also sub-phrases that reflect the desired sentiment and style. 

% removed for now b/c paper too long
% \header{Adjusting temperature}
% % By adjusting the temperature parameter of a model's output softmax distribution, 
% One can adjust the ``peakiness'' of a model's output softmax distribution through the temperature parameter to result in more conventional or random samples. Similar to behaviors seen in prior work~\cite{louie2020novice,sturm2019machine}, we observed participants using temperature as a proxy to control certain musical attributes: \textit{``If I wanted a slow paced sequence, I would reduce the temperature''} (T12).
%
%  adjusts the parameter more
% formally known as the temperature (T) of the sampling distribution [20]. A lower temperature makes the distribution more
% “peaky” and even more likely for notes to be sampled that had
% higher probabilities in the original distribution (conventional),
% while higher temperatures makes the distribution less “peaky”
% and sampling more random (surprising). 
%

% \header{Adjusting the temporal resolution}
% Generative music models are often trained on input with a predetermined temporal resolution. By adjusting the resolution of new input, the model can ``listen'' to it in a different spend. For example, one team member reported that \textit{``If I wanted short timed notes, I would increase the steps per quarter used during generation''} (T12). 

%\todo{is this during training or during generation? I don't remember coconet or musicvae having these properties at generation time...}
% it's a data preprocessing hack by stretching or compressing a sequence before feeding it into the model

\header{Priming} While co-creating, teams often desired to create musical content in a particular key, chord, contour, or pitch range. To do so, many attempted to reverse engineer the process by priming the model's input sequence using music containing the desired property, in the hopes that it would produce a continuation with similar characteristics: \textit{``I found that I could control the generation process by using different kinds of input. So if I wanted a sequence that is going upward, I would give that kind of input. If I wanted the model to play certain notes, I would add those notes in the input pattern''} (T12). This seemingly simple way of requesting for continuations led to a wide range of controls when used creatively.
%, which was a successful strategy for this team: \emph{``Since there are so many options available, there is an opportunity to be creative with it in a definite way, and that works best for me!''}

To further direct content in lyrics, a team used specific words to start off each sentence (T5). Another team wondered \textit{``can \eurovision songs convey insightful messages with only two words?''}, and put together a verse with frequent bigrams from that dataset, such as \textit{``my love''}, \textit{``your heart''}. They entered this verse through the TalkToTransformer~\cite{king2019talk} web-app interface as context for GPT-2 to generate the next verses (T11).

%%\todo{Similarly, another team primed GPT-2 with \emph{``I rock the world''} as the first line of their lyrics because each of the words were highly frequent in English lyric datasets, also because \emph{``it's really funny in French''} (T4).}

%first sentence of our lyrics “I rock the world” based on the words frequency of your english lyrics dataset

%During generation time, a model's output can be directed by priming it with an initial input sequence.
%Priming allowed the \participantt to play ping-pong with the \mll model, steerin its continuations towards his desired key, chords, contour, note patterns and pitches.
%  given models an initial sequence, and then requesting continuations (i.e. beginning of lines in lyrics, seed melody in a certain key)

% TODO: could save this for the overcoming limitation section as examples of steering a model

% Controlling/steering/directing what the models generate [analogously, model generated material can direct what is manually composed too!]
% Unconditioned generation from model that best aligns with desired musical style
%
% \header{Primed generation} given models an initial sequence, and then requesting continuations (i.e. beginning of lines in lyrics, seed melody in a certain key)
% Iterating / ping pong between human and AI

\header{Interpolating} Models such as MusicVAE provide a continuous latent space that supports interpolating between existing sequences to generate new sequences that bear features of both~\cite{roberts2018learning}. Some teams leveraged this latent space as a way to explore. For example, one participant found by chance that \textit{``interpolating between really simple sequences at high temperatures would end up giving me these really cool baseline sequences''} (T12). 
%\todo{this seems more like a technique for exploring rather than trying to steer things purposefully towards a desired property. did anyone mention using this in a more goal-directed way?}

% the example by T13 where they use coconet on a rave melody is a good example?
%%This opened up a rich space for exploration, and 
% covered in first section

%
% \header{Curating generated output}
% How to decide which sample to select?
% Automate selection through critic model (i.e. “catchiness”)
% Often by combining multiple musical components together 
%
% \header{Cascaded generation}
% Cascade/pipeline generation: a sequence of conditioned generation on different musical components (sometimes interleaving manual and AI?)
%\vspace{-3pt}
\subsection{\mll is not structure-aware}
Because large, end-to-end models were not easily decomposable into meaningful musical structures, most teams %resorted to composing using 
used multiple smaller, independent models. Yet, these smaller, independent \mll models are not able to take into account the holistic context of the song when generating local snippets. To address this, users created their own workarounds to fill in this contextual gap, by arranging and weaving these independent pieces into a coherent whole.
%in addition to taking a modular approach with respect to musical components, \teams also took a modular approach to building out the structure of a piece in the co-creation process.

\header{Creating an overall song structure} To create a backbone for the song, some teams used their musical knowledge to first curate chord progressions that can serve well for each song section (i.e. verse, chorus). One team then used a conditioned model (pretrained to be conditioned on chord progressions) to generate melody and basslines that would go well with those chords (T3). 

%\header{Creating an overall song structure} To create a backbone for the song, some teams used their own knowledge of chord progressions as a way to carve out an overall trajectory for the song. For example, a team first generated and curated chord progressions that they felt fitting for each section of the song. They then used a conditioned model (pre-trained to be conditioned on chord progressions), providing it a chord conditioning signal to generate a melody and bassline that would go well with those chords (T3). 
%The team was extremely surprised and happy when first hearing the output: \emph{``The architecture seemed to produce results we could be happy with and listen to even in our free time.''} However, the team felt like they might have over-complicated the design of their model, as they found some services that provide comparable results with a plug an play solution (T3). 

\header{Creating contrast between sections}
The verse and chorus sections of a song often carry contrast. However, participants did not have a direct way to express this desire for structured contrast while preserving overall coherence between verse and chorus. To address this, one team used their verse as a starting point to generate various continuations to the melody and bass lines, and finally chose a variation for the chorus section (T2). 
% choosing the ones that fit to be the lead for the chorus section (T2).
% \todo{I don't understand what ``chose the ones that fit to be the lead" means}. 
%Another team used similar priming approaches to generate both the verse and chorus simultaneously, but used different temperatures in order to \textit{``add some randomness into the generation''} (T12). These approaches gave users a way to manually create structured variety, through stitching sections together.
Another team used similar priming approaches but used different temperatures to generate the verse and chorus in order to \textit{``add some randomness into the generation''} (T12). These approaches gave users a way to manually create structured variety.%, through stitching sections together.

% Creating musical contrast/variation to build structure/progression in song
% ​From this 16 bar-track, we handpicked the most
%  appealing part as a verse​. The chorus and bridge were created in Magenta Studio: we used it to
%  create basslines and melodies originating from our first 16-bar track. Other parts of the song
%  were produced in a similar fashion.

% ``Each section can be sampled at different temperatures (which adds randomness into the generation)'' (T12).
% Adjusting the temperature of models when generating to adjust the character

% Finetuning on different small datasets for different sections

% \todo{what was the outcome? were the different parts interesting in combination because the underlying chord conditioning was seeded by this context-awareness? or did that not work very well either}
%In contrast to priming a model with notes from a desired chords, %A team first generated and curated chord progressions they felt fitting for each section, and then used a chord conditioned model to generate the melody and bass line 
% Conditioned generation: conditioning on another modality/music component (i.e. condition on chords to generate melody and baseline)

%One interesting strategy for preserving holistic structure 
\header{Rewriting to add variation} Rewriting allows one to generate new material while borrowing some structure from the original material. For example, one team was able to generate a \textit{``darker version''} of the chorus of another song by rewriting it multiple times, alternating between re-generating the melody conditioned on the accompaniment, and then re-generating the accompaniment conditioned on the melody. To create a coherent song structure, the team initially attempted to \textit{``repeat the same rave section twice"}, but later \textit{``realized that was boring''}. 
The team then decided to vary how the second section ended by reharmonizing the melody with a new flavor: \textit{``Coconet is great at reharmonizing melodies with a baroque flavor. We entered in the notes from our rave chorus. After a few tries, it made this awesome extended cadence''} (T13).

\header{Bridging sections through latents}
% For models with latent spaces such as MusicVAE~\cite{roberts2017hierarchical}, the MidiMe~\cite{dinculescu2019midime} approached show that mini MusicVAEs can be trained on a set of melodies or trios to generate variations in that style. 
One team devised an unusual strategy for connecting different sections of a song in a meaningful way (T9). They first trained multiple miniVAEs~\cite{dinculescu2019midime}, one for each section of the song (e.g. intro, verse, chorus or genre such as rock and pop). They then composed the song by computing the \emph{``shortest path''} through these latent spaces, allowing the sections to share elements in common with each other, despite each being distinctive. The genre miniVAEs also made style transfer possible by interpolating an existing trio towards their latent areas, allowing them to tweak the style of each section. 
%\todo{last sentence: need to explain this a bit more in detail, feels a bit cursory}

%\vspace{-3pt}
\subsection{ML setup can interfere with musical goals}
A logistical hurdle faced by teams was the significant setup and customization issues encountered to even start composing with a model. 
% \todo{and software compatibility (T4)} 
Aside from musical considerations, many teams chose or abandoned models based on what was available out-of-the-box, such as pre-trained checkpoints, fine-tuning scripts, scripts for retraining on a different dataset, data pre-processing scripts. In addition, different models expected different types of music representation as input (e.g. MIDI, piano roll, custom), adding additional pre-processing overhead to begin experimenting with them. To decrease time spent model wrangling, large teams sometimes divide-and-conquered, exploring several models in parallel to see which worked best. Ultimately, teams' co-creation process involved navigating not only their musical goals, but also the logistical costs and benefits of using one model or another.

%\todo{Not sure if this is completely in line with the paragraph:}
%For some teams, ML interfered with their (inital) musical goals. Examples are teams who aimed for end-to-end audio generation, but were not able to generate acceptable results and changed their strategies to modular approaches (T13,T6). Another team aimed for using MusicVAE, but was forced to change their approach to a different model because of an incompatible Tensorflow update (T4). Because of time constraints, one team could not create a more complicated model incorporating chord generation based on time-based melody matching (T5).

%\todo{Name an example where ML setup constraints interfered with a musical goal. Name a few strategies teams adopted to short-circuit all this overhead, if any} 

%creative prototyping” with ML models 
%\vspace{-5pt}
\section{Discussion}

\subsection{Decomposeable and context-aware modeling}
% \subsection{Support for more structural rapid prototyping in co-creation} 
Writing a song involves composing multiple sections, and multiple vocal and instrumental parts that all need to work well together as a whole. Because end-to-end models lacked meaningful hierarchical structures, and because smaller models lacked global awareness, participants often needed to reverse engineer a multitude of models, applying heuristics and domain knowledge to approximate high-level structure or to achieve desired musical effects.  
%To do this, musicians applied their deep understanding of song structure, such as creating two versions of a chorus that vary in their ending. Yet, b
%How can we get our building block models to jam together better? How can we decompose end-to-end models so that we can steer them more directly? Given a verse as context, can a model better prototype a chorus, to create the needed contrast and variation that works for a song?
To ameliorate this process, one approach could be to infuse smaller models with more context-awareness, and exposing the common ways that they can be customized through user-facing controls. For example, a melody model could allow users to express whether they are creating a verse as opposed to a chorus, or whether they would like it to contrast with the next section. Another possibility is to design end-to-end models to have intermediate representations that align with the musical objects that musicians already know how to work with. The sweet spot is probably a hybrid that combines the flexibility of smaller models with the benefits of global context in end-to-end modeling.

\subsection{AI-defined vs. User-defined building blocks}
The design of \mll models for music involves a series of upstream decisions that can have a large impact on how musicians think about music when they co-create with these models downstream. Whereas regular songwriting often starts with an initial spark of a human idea, in this work we found that the practical availability and limitations of AI tools were instead key drivers of the creative process, defining the scope of what's musically legitimate or possible. For example, most teams broke down songwriting into lyrics, melody, chords, etc., in part because these models were readily available. Yet, there are also other music building blocks that do not have corresponding, ready-made generative modsels (e.g. motif, verse, chorus, coda, etc.) or that currently are not treated as separate, first-class building blocks in deep models (e.g. rhythm, pitch). Likewise, a musician's creative thought process can be unintentionally influenced by the order of steps taken to fine-tune, condition, prime, and apply a model. In the future, the design of \mll models should be coupled with a more careful consideration of what workflows and building blocks end-users already use in their existing practice, and perhaps start with those as first-class principles in guiding the design of AI systems.

\subsection{Support for parallel music and ML exploration}
\begin{figure}[h!]
 \centerline{
 \includegraphics[width=0.95\columnwidth, trim={0cm 0cm 0cm 0cm},clip]{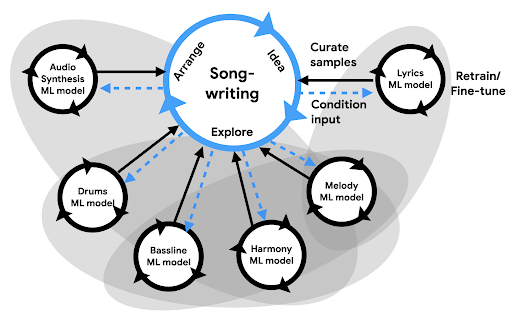}}
 \caption{Parallel music and ML feedback loops in human-AI co-creative songwriting.}
 \label{fig:parallel}
\end{figure}

A central aspect of the creative process involves a ``flare and focus"~\cite{buxton2010sketching} cycle of ideating, exploring those ideas, selecting ideas, then rapidly iterating. We found that a key challenge of human-AI co-creation was the need to juggle not only this \textit{creative} process, but also the \textit{technological} processes imposed by the idiosyncrasies and lack of steerability of learning algorithms. For instance, while ideating motifs for a song, participants needed to carry out a large additional burden of sub-tasks, such as selecting which combination of models to use, re-training or conditioning them as necessary, chaining them together, and ``gluing" their outputs together. In essence, the typical ``flare and focus" cycles of creativity were compounded with a parallel cycle of having to first explore and curate a wide range of models and model outputs (Figure~\ref{fig:parallel}). While some of these model-wrangling tasks led to new inspiration, many interfered with the rapid-iteration cycle of creativity. %need for heavy-handed model exploration and wrangling also meant that initial model-based decisions sometimes prematurely constrained the subspace of what could musically be done downstream.
%\todo{In practice, these technical challenges and overhead causes a division of labor, where developers are tasked to setting up a usable ML system to iterate with, which can cause musicians to be left outside of the ML-loop where the exploration and curation of musical material happens.} 

%trim={1.2cm 0.8cm 4.0cm 1.5cm}

These issues raise important questions around how best to support users in juggling the dual processes of creative and technological iteration cycles. One approach is to have ML models readily available to musicians in their natural workflows. For example, Magenta Studio~\cite{roberts2019magenta} makes available a suite of model plug-ins to music production software such as Ableton Live~\cite{ableton}, and Cococo~\cite{louie2020novice} allows users to semantically steer a model directly in its user interface. Beyond this, human-AI interfaces could scaffold the \textit{strategic} part of the model exploration and selection process 
% by surfacing commonly used model combinations (e.g. melody and \todo{X and Y}) 
by surfacing effective model combinations (e.g. using general infilling models for rewriting or to reharmonize another generated melody) 
%Coconet or MusicAutobot 
or fruitful workflows (e.g. matching lyric and melody stress patterns), so that new users can benefit from past users' experiences. Reducing this overhead of model-based decisions could empower users to more easily prototype their creative ideas, accelerating the feedback and ideation cycle.

\vspace{-9pt}
\section{Conclusion}
We conducted an in-depth examination of how people leverage modern-day deep generative models to co-create songs with AI. We found that participants leveraged a wide range of workarounds and strategies to steer and assemble a conglomeration of models towards their creative goals. These findings have important implications for how human-AI systems can be better designed to support complex co-creation tasks like songwriting, paving the way towards more fruitful human-AI partnerships.

\section{Acknowledgements}
We thank Karen van Dijk and VPRO for organizing the AI Song Contest, and also NPO Innovatie, 3FM and EBU. We want to thank all participants for their insights and contributions. 
We also thank Tim Cooijmans for creating early versions of the figures and Michael Terry for feedback on this manuscript.

\bibliography{aiSongContest}

\end{document}